4/IV/2014; **Strange magnetic multipoles and neutron diffraction by an iridate perovskite ($Sr_2IrO_4$)**

S W Lovesey[1,2], and D D Khalyavin[1]

1. ISIS Facility, STFC Oxfordshire OX11 0QX, UK

2. Diamond Light Source Ltd, Oxfordshire OX11 0DE, UK

**Abstract**

A theoretical investigation of a plausible construct for electronic structure in iridate perovskites demonstrates the existence of magnetic multipoles hitherto not identified. The strange multipoles, which are parity-even, time-odd and even rank tensors, are absent from the so-called $j_{eff}$ = 1/2 model. We prove that the strange multipoles contribute to magnetic neutron diffraction, and we estimate their contribution to intensities of Bragg spots for $Sr_2IrO_4$. The construct encompasses the $j_{eff}$ = 1/2 model, and it is consistent with the known magnetic structure, ordered magnetic moment, and published resonant x-ray Bragg diffraction data. Over and above time-odd quadrupoles and hexadecapoles, whose contribution changes neutron Bragg intensities by an order of magnitude, according to our estimates, are relatively small triakontadipoles recently proposed as the primary magnetic order-parameter of $Sr_2IrO_4$.

**1. Introduction** By and large, recent attempts to interpret and unify properties of iridate perovskites utilize a one-electron, spin-orbit-entangled ground-state derived from the $t_{2g}$ manifold of $Ir^{4+}$ ($5d^5$) orbitals. Numerous simulations of magnetic properties provide a sound basis for such a candidate, and a nexus with quantum phenomena that include spin-liquid behaviour, quantum spin Hall effect, quantum compass, and a Kitaev model [1 - 10]. Structure within a $t_{2g}$ ground-state descends from relative magnitudes of three interactions; super-exchange, tetragonal distortion of the crystal-field potential, and spin-orbit coupling, listed with ascending orders of magnitude. The branching ratio observed in x-ray dichroism is a notable failure of a one-electron ground-state, however. Measured values are reproduced, to a good approximation, by full multi-electron simulations with $t_{2g}$ and $e_g$ hole states and a subservient tetragonal distortion, while a one-electron $t_{2g}$ model falls far short [11 - 15].

Current questions about magnetic properties of $Sr_2IrO_4$ include; (i) is the so-called $j_{eff}$ = 1/2 model of real value? [1, 14, 15] and (ii) what is the magnetic order-parameter? [10]. In this communication we propose magnetic neutron diffraction experiments that can shed light on both questions, with feasibility already established [16, 17]. At the same time, we introduce entities that are likely new in the theory of magnetic neutron scattering. Whereas magnetic multipoles usually encountered have an odd tensor rank, e.g., a magnetic dipole, we encounter magnetic (time-odd) multipoles with even rank.

(i) The conventional definition of the $j_{eff}$ = 1/2 model adopts exclusive use of the J = 5/2 electron manifold of $Ir^{4+}$ ($5d^5$) orbitals [1, 5, 10, 18, 19, 20]. Absence of J = 3/2 is counter-intuitive in so far as the established distortion of the environment from cubic (tetragonal) symmetry appears to be an important attribute, and the distortion mixes states with J = 3/2 and J = 5/2 [12, 14]. Strange time-odd multipoles with even rank, constructed from the spin anapole [22. 23], and not previously been discussed, to the best of our knowledge, are signatures of a mixed J-state. The time-odd and parity-even quadrupole and hexadecapole are visible in magnetic Bragg diffraction of neutrons by an iridate, given a spin-orbit-entangled ground state.

(ii) Ganguly et al [10] simulate magnetic properties of $Sr_2IrO_4$ adopting a space-group $P_Icca$ proposed by Lovesey et al [21]. They present arguments in favour of a primary magnetic order-parameter proportional to an $Ir^{4+}$ atomic multipole of rank 5 (triakontadipole), and support a $j_{eff}$ = 1/2 model. The dipole order-parameter in this scenario is a secondary order-parameter coupled to the primary through a bilinear free-energy invariant. Triakontadipoles are shown by us to contribute to the magnetic neutron Bragg diffraction amplitude, rendering them open to experimental scrutiny.

By way of orientation to what follows in the main text, we recount fundamental properties of magnetic neutron scattering and associated electronic multipoles. The magnetic neutron-electron interaction, **Q**, defined in (2.1) is a dipole with discrete symmetries identical to those of the magnetic moment, i.e., **Q** is parity-even and time-odd. For the moment we need only observe that, without loss of generality, **Q** can be expanded in terms of neutron and electron operators, $V^K(n)$ and $V^{K'}(e)$, respectively, where positive integers K and K' are ranks of the spherical tensors. A tensor product $\{V^K(n) \otimes V^{K'}(e)\}^1$ couples the operators to form a dipole and,

$$\mathbf{Q} = \Sigma_{KK'} \{V^K(n) \otimes V^{K'}(e)\}^1. \qquad (1.1)$$

Allowed values of K and K' are fixed by physical properties of **Q** and properties of unpaired electrons in the material of interest. They obey the triangle rule for a dipole, of course, with K = K' − 1, K', K' + 1.

In many cases, symmetry in the electronic structure limits K' to odd integers; more on this later. A useful dipole approximation to **Q** is derived from K' = 1. Octupole (K' = 3), triakontadipole (K' = 5) and higher-order electronic contributions are discarded in this approximation on the grounds that (a) electronic multipoles with K' > 1 are absent at the forward direction of scattering (scattering wavevector **k** = 0) when **Q** is proportional to the magnetic moment of an ion and (b) maximum magnitudes of multipoles diminish with increasing K'. The dipole approximation is adequate if the goal is nothing more than determination of the motif of magnetic moments. However, if the brief for an experiment goes beyond the motif, and includes finding detailed information on electronic degrees of freedom, then all multipoles allowed by symmetry must be included in the analysis of data.

Operators $V^K(n)$ and $V^{K'}(e)$ have discrete symmetries such that their product in (1.1) has symmetries of a magnetic moment - to match those of **Q**. It is standard practice to choose $V^K(n)$ to be a function of the direction of the scattering wavevector, **κ** = **k**/k. In which case, $V^K(n)$ and $V^{K'}(e)$ are time-even and time-odd, respectively. The neutron operator is parity-even (parity-odd) for K even (K odd). And the parity of $V^{K'}(e)$ is fixed by the unpaired electrons. If unpaired electrons engaged in scattering are in an environment that possesses a centre of inversion symmetry $V^{K'}(e)$ is parity-even. In the absence of a centre of inversion symmetry, $V^{K'}(e)$ can be parity-odd and K odd. Dipoles that contribute to $V^{K'}(e)$ in this case include a spin anapole (**s** x **n**), where **s** and **n**, respectively, are electron spin and electric dipole operators. The spin anapole is manifestly time-odd and parity-odd, and the corresponding $V^K(n)$ is the dipole **κ**, i.e., K = K' = 1. A dipole approximation to **Q** contains the operator i[**κ** x (**s** x **n**)], and a similar entity constructed from orbital angular momentum.

Even values of K' are allowed whether or not there is a centre of inversion symmetry. They are forbidden only when unpaired electrons belong to a state with both unique orbital angular momentum and unique total angular momentum. In consequence, multipoles with K' even are candidates in a spin-orbit-entangled state of the type proposed for iridate perovskites, which neglects inversion breaking [21]. Multipoles in question are a product of (**s** × **n**) with odd multiples of **n** that together form a multipole that is parity-even.

Main results in this communication are magnetic unit-cell structure factors for $Sr_2IrO_4$ derived from the established magnetic motif, depicted in Figure 1 (the magnetic structure is orthorhombic, with magnetic crystal-class mmm1', whereas the chemical structure is tetragonal, $I4_1/acd$, crystal class 4/mmm) [21]. All electronic multipoles encountered in our unit-cell structure factors can be estimated with the approach initiated by Ganguly et al [10]. Unit-cell structure factors are reported in the next section, and estimates of multipoles therein appear in Section 3. Section 4 is given over to multipoles in neutron diffraction, with attention to time-odd, parity-even multipoles with even rank. Discussion and conclusions appear in Section 5.

**2. Simulation of magnetic neutron diffraction by $Sr_2IrO_4$** A magnetic interaction operator for neutrons, $Q_\perp$, is the sum of electron spin and linear momentum field operators [24, 25, 26]. There is advantage in using an auxiliary operator,

$$\mathbf{Q} = \Sigma \exp(i\mathbf{R}\cdot\mathbf{k}) \, [\mathbf{s} - (\mathbf{k} \times \nabla)/k^2], \qquad (2.1)$$

with $\mathbf{Q}_\perp = [\mathbf{k} \times (\mathbf{Q} \times \mathbf{k})]/k^2 = \mathbf{Q} - \kappa\,(\kappa\cdot\mathbf{Q})$. In (2.1), the sum is over all unpaired electrons in the material of interest, **k** is the scattering wavevector, and **R**, **s**, and $\mathbf{p} = -i\hbar\nabla$ are, respectively, position, spin and linear momentum operators.

The basis of a simulation of Bragg diffraction by $Sr_2IrO_4$ is an electronic structure factor for the motif of magnetic dipoles depicted in Figure 1, which belongs to irreducible representation $M_4$ in the magnetic space-group $P_Icca$ [21]. Throughout the following discussion, we reserve K for even rank, and K' for odd rank spherical multipoles, $\langle T^{K'}_{Q'} \rangle$ and $\langle T^K_{Q'} \rangle$, where $\langle ... \rangle$ denotes the time-average of the enclosed operator. The point group for iridium ions, $C'_{2c}$, restricts projections Q' to odd integers for both K and K'. Electronic structure-factors, $\Psi^{K'}_{Q'}$ and $\Psi^{K}_{Q'}$ for Bragg diffraction are constructed with $\langle T^{K'}_{Q'} \rangle$ and $\langle T^{K}_{Q'} \rangle$ using elements of symmetry in $P_Icca$.

Space-group forbidden Bragg spots $\mathbf{k} = (h, 0, l)$ and $(0, k, l)$, with $l$ even and $h$ and $k$ odd, possess $\Psi^{K}_{Q'} = \Psi^{K}_{-Q'}$ and $\Psi^{K'}_{Q'} = -\Psi^{K'}_{-Q'}$. These identities determine the corresponding magnetic amplitude for Bragg diffraction, $\langle \mathbf{Q}_\perp \rangle$. Let us continue with $\mathbf{k} = (\sin\theta_o, 0, \cos\theta_o) |(h, 0, l)|$, and in subsequent work use a shorthand $c = \cos\theta_o$ and $s = \sin\theta_o$. We find $\langle Q_y \rangle \equiv \langle Q_b \rangle = 0$, and $\langle \mathbf{Q}_\perp \rangle = F(\mathbf{k}) (-c, 0, s)$, with a purely real unit-cell structure factor,

$$F(\mathbf{k}) = (s \langle Q_z \rangle - c \langle Q_x \rangle) = c\, [(3/\sqrt{2})\, \Psi^1_{+1} + (\sqrt{21}/8)\, (15\, c^2 - 11)\, \Psi^3_{+1}$$

$$+ (3\sqrt{35}/8)\, s^2\, \Psi^3_{+3} + (\sqrt{11}/8\sqrt{2})\, (105\, c^4 - 126\, c^2 + 29)\, \Psi^5_{+1}$$

$$+ (\sqrt{231}/16)\, s^2\, (15\, c^2 - 7)\, \Psi^5_{+3} + (\sqrt{1155}/16)\, s^4\, \Psi^5_{+5} \qquad (2.2)$$

$$+ \sqrt{3}\, \Psi^2_{+1} + (\sqrt{3}/4)\, (7\, c^2 - 3)\, \Psi^4_{+1} + (3\sqrt{21}/4)\, s^2\, \Psi^4_{+3}].$$

Electronic structure-factors appearing in $F(\mathbf{k})$ for Bragg spots $(h, 0, l)$ with $h$ odd and $l = 4n$ are [21],

$$\Psi^{K'}_{Q'} = -8\, (-1)^n\, \langle T^{K'}_{Q'} \rangle' \text{ with } K' \text{ odd,}$$

$$\Psi^{K}_{Q'} = 8\, (-1)^n\, \langle T^{K}_{Q'} \rangle'' \text{ with } K \text{ even,} \qquad (2.3)$$

where $Q'$ is odd, and ' (") denotes the real (imaginary) part of a spherical multipole. In the forward direction, $\mathbf{k} \Rightarrow 0$, $\Psi^1_{+1} \Rightarrow (4\sqrt{2}/3)\, (-1)^n\, \langle \mathbf{l} + 2\mathbf{s} \rangle_x$. On the other hand, for $(0, 2m + 1, 4n + 2)$ electronic structure factors in $F(\mathbf{k})$ are,

$$\Psi^{K'}_{Q'} = -8\, (-1)^{m+n}\, \langle T^{K'}_{Q'} \rangle' \text{ with } K' \text{ odd,}$$

$$\Psi^{K}_{Q'} = 8\, (-1)^{m+n}\, \langle T^{K}_{Q'} \rangle'' \text{ with } K \text{ even.} \qquad (2.4)$$

The Bragg spot $(0, 1, 2)$ is the strongest magnetic reflection, because of the relatively small magnitude of the Bragg wavevector and favourable orientation of the magnetic moment [21].

A notable feature of the unit-cell structure factor (2.2) is the appearance of tensors of even rank trailered in Section 1. Regarding point (i), magnetic quadrupoles and hexadecapoles are zero if the Ir ground-state is pure $J = 5/2$. For point (ii) we note in $F(\mathbf{k})$ the presence of tensors of rank $K' = 5$.

We move to a discussion of Bragg spots (0, 0, l) with l odd, for which $\Psi^K_{Q'} = - \Psi^K_{-Q'}$ and $\Psi^{K'}_{Q'} = \Psi^{K'}_{-Q'}$, with Q' odd. In consequence, intensity of Bragg spots (0, 0, l) with l odd uses the weak moment parallel to the b-axis, whereas intensities of Bragg spots (h, 0, l) and (0, k, l) with l even and h and k odd use the moment parallel to the a-axis [12]. Dependence of $\langle \mathbf{Q}_\perp \rangle$ on **k** arises only from radial integrals, included in our multipoles, and there is no angular dependence unlike the corresponding $\langle \mathbf{Q}_\perp \rangle$ for (h, 0, l) and (0, k, l). In addition to $\langle \mathbf{Q}_{\perp,z} \rangle \equiv \langle \mathbf{Q}_{\perp,c} \rangle = 0$, a calculation reveals $\langle \mathbf{Q}_{\perp,x} \rangle \equiv \langle \mathbf{Q}_{\perp,a} \rangle = 0$ and $\langle \mathbf{Q}_{\perp,y} \rangle \equiv \langle \mathbf{Q}_{\perp,b} \rangle = i\sqrt{2}\langle \mathbf{Q}_{\perp,+1} \rangle$ with,

$$\langle \mathbf{Q}_{\perp,+1} \rangle = \langle \mathbf{Q}_{\perp,-1} \rangle = \sqrt{(3/2)} \, [\Psi^2_{+1} + \Psi^4_{+1}]$$

$$+ \Sigma_{K'} [3(2K' + 1)/2(K' + 1)]^{1/2} \Psi^{K'}_{+1}, \quad (2.5)$$

where K' = 1, 3, 5. Note that only one projection is visible, |Q'| = 1. Electronic structure factors are (Q' is odd) [21],

$$\Psi^{K'}_{Q'} = i \, 8 \cos(\pi l/4) \, \langle T^{K'}_{Q'} \rangle'' \text{ with K' odd,}$$

$$\Psi^K_{Q'} = i \, 8 \cos(\pi l/4) \, \langle T^K_{Q'} \rangle' \text{ with K even,} \quad (2.6)$$

and $\langle \mathbf{Q}_{\perp,y} \rangle$ is purely real. In particular, $\langle \mathbf{Q}_{\perp,y} \rangle = 4 \cos(\pi l/4) \langle \mathbf{l} + 2\mathbf{s} \rangle_y$ in the forward direction **k** = 0.

**3. Estimates of multipoles** Let us assess multipoles with the model wavefunction deployed by Chapon and Lovesey [12], derived from the $t_{2g}$ manifold in the presence of a tetragonal crystal potential, with axial distortion along the c-axis, and spin-orbit coupling. In this case, the $t_{2g}$ manifold is usually mapped to an effective orbital angular momentum $l_{eff}$ = 1, whereupon a ground-state is a mixture of spin-orbital product states $|m_{eff} = +1, \downarrow\rangle$ and $|m_{eff} = 0, \uparrow\rangle$ (the state $f_\uparrow$ in reference [7]). The mixing angle, δ, is related to the ratio of the spin-orbit coupling parameter to the axial distortion, and $\cos\delta = \sqrt{2} \sin\delta = \sqrt{(2/3)}$ for a cubic environment. In terms of states $|J, M\rangle$ a suitable component of the ground-state wavefunction is,

$$|u\rangle = \alpha|3/2, -3/2\rangle + \beta|5/2, -3/2\rangle + \gamma|5/2, 5/2\rangle, \quad (3.1)$$

where $\alpha = (1/\sqrt{5})(-\cos\delta + \sqrt{2}\sin\delta)$, $\beta = (1/\sqrt{5})(2\cos\delta + \sin\delta/\sqrt{2})$, and $\gamma = -(\sin\delta/\sqrt{2})$. Note that in the absence of a distortion α = 0, and the remaining pure

J = 5/2 state is the basis of the conventional $j_{eff}$ = 1/2 model [1, 5, 10, 18, 19, 20].

A ground-state wavefunction, $|g\rangle$, is a linear combination of $|u\rangle$ and the conjugate state of the Kramers doublet $|\hat{u}\rangle$. With $|g\rangle = (|u\rangle + f|\hat{u}\rangle)/\sqrt{2}$ and $|f|^2 = 1$ magnetic dipoles lie in the a-b plane, in accord with the established magnetic motif. The corresponding saturation magnetic moment $\mu = \sqrt{\{\mu^2_a + \mu^2_b\}} = |\sin\delta (\sin\delta + \sqrt{2} \cos\delta)|$ is independent of f. If $\phi$ is the angle enclosed by the moment and the crystal a-axis [100] we find $f = \exp(i\phi)$. The ground-state $|g\rangle$ is consistent with resonant Bragg diffraction of x-rays, in that there is no intensity at the Ir $L_2$ absorption edge [12, 18]. While we submit that our ground-state wavefunction is plausible its shortcomings include a wrong branching ratio and no allowance for quantum fluctuations that likely influence the ordered magnetic moment.

In expressions for magnetic multipoles, we use a notation $\langle g|...|g\rangle = \langle...\rangle$, and find;

$\langle T^1_{+1}\rangle = 0.211\ f\ \gamma\ [\alpha\ (-\langle j_0\rangle + (1/2)\langle j_2\rangle) - 3\beta\ (\langle j_0\rangle + (4/7)\langle j_2\rangle)]$,

$\langle T^3_{+1}\rangle = 0.028\ f\ \gamma\ [\alpha\ (-\langle j_2\rangle + (3/4)\langle j_4\rangle) - 2\beta\ (9\langle j_2\rangle + 2\langle j_4\rangle)]$,

$\langle T^3_{+3}\rangle = 0.048\ f^*\ [\alpha^2\ (\langle j_2\rangle + 8\langle j_4\rangle) - 2\beta^2\ (3\langle j_2\rangle + (2/3)\langle j_4\rangle)$

$\qquad\qquad + \alpha\beta\ (\langle j_2\rangle - (3/4)\langle j_4\rangle)]$,  (3.2)

$\langle T^5_{+1}\rangle = -0.136\ f\ \beta\gamma\ \langle j_4\rangle$,  $\langle T^5_{+3}\rangle = 0.329\ f^*\ \beta^2\ \langle j_4\rangle$,

$\langle T^5_{+5}\rangle = -0.441\ f\ \gamma^2\ \langle j_4\rangle$,

$\langle T^2_{+1}\rangle = i\ 0.461\ f\ \alpha\gamma\ \langle j_2\rangle$,  $\langle T^4_{+1}\rangle = i\ 0.138 f\ \alpha\gamma\ \langle j_4\rangle$,

$\langle T^4_{+3}\rangle = i\ 0.818\ f^*\ \alpha\beta\ \langle j_4\rangle$.

Here, integer values have been converted to decimal values, e.g., $(2/7\sqrt{105}) \approx 0.028$. Radial integrals $\langle j_n\rangle$ with n even are normalized such that $\langle j_0\rangle = 1$, and all other $\langle j_n\rangle = 0$, for k = 0 [28].

Triakontadipoles do not depend on $\alpha$, whereas even rank multipoles are proportional to $\alpha$ and absent in the $j_{eff}$ = 1/2 model, as anticipated. The quadrupole and hexadecapoles and three triakontadipoles arise solely from the operator $\exp(i\mathbf{R}\cdot\mathbf{k})\mathbf{s}$. Dipoles and octupoles contain additional contributions created by $\exp(i\mathbf{R}\cdot\mathbf{k})$ ($\mathbf{k} \times \nabla$) that account for the additional complexity of expressions in (3.2). Dipole moments have values,

$$3 \langle T^1_x \rangle \Rightarrow - \cos(\phi) \sin\delta (\sin\delta + \sqrt{2} \cos\delta) = \mu_x,$$

$$3 \langle T^1_y \rangle \Rightarrow - \sin(\phi) \sin\delta (\sin\delta + \sqrt{2} \cos\delta) = \mu_y, \qquad (3.3)$$

for $\mathbf{k} \Rightarrow 0$. All other multipole moments vanish in the forward direction.

A fair estimate of the ordered magnetic moment required, since results (3.3) remind us that $\mu = \sqrt{\{\mu^2_a + \mu^2_b\}}$ sets the scale of Bragg intensities. From four values of $\delta$ that satisfy $\mu = 0.21 \mu_B$ [16] there is one that corresponds to a spin-orbit parameter twice as large as the tetragonal distortion, and oxygen octahedra elongated along the c-axis. With $\delta = 135°$ product states $|m_{eff} = + 1, \downarrow\rangle$ and $|m_{eff} = 0, \uparrow\rangle$ have equal weight in the ground-state. Corresponding coefficients in $|u\rangle$ are $\alpha = 0.76$, $\beta = - 0.41$ and $\gamma = - 0.50$.

Figure 2 shows our estimates of intensity in Bragg spots (1, 0, 4n). In one calculation all multipoles in (2.2) are included in evaluation of $\{F(\mathbf{k})\}^2$ and in a second calculation the quadrupole and hexadecapoles are omitted. Omission of the quadrupole and hexadecapoles, multipoles that epitomize the $j_{eff} = 1/2$ model, causes a stunning increase in $\{F(\mathbf{k})\}^2$. On the other hand, triakontadipoles (rank 5) make a small contribution to intensities because they are proportional to $\langle j_4 \rangle$. The shape of $\{F(\mathbf{k})\}^2$ as a function of $\mathbf{k}$ reflects radial integrals $\langle j_2 \rangle$ and $\langle j_4 \rangle$ whose contribution is exaggerated by a small value of the magnetic moment (a similar form factor is exhibited by samarium, where also the magnetic moment is small because spin and orbital dipoles have opposite signs). Radial integrals are taken from reference [28], and displayed in Figure 2.

To assess our startling results for $\{F(\mathbf{k})\}^2$ in Figure 2 let us look at its value at (1, 0, 4). Radial integrals $\langle j_0 \rangle = 0.768$, $\langle j_2 \rangle = 0.088$ and $\langle j_4 \rangle/\langle j_2 \rangle = 0.041$ imply that $\langle T^1_{+1} \rangle$ and $\langle T^2_{+1} \rangle$ are the most significant multipoles at the chosen small wavevector. With $\langle T^1_{+1} \rangle' = - 0.0467$ and $\langle T^2_{+1} \rangle'' = - 0.0149$ we find $\{(3c/\sqrt{2}) \Psi^1_{+1}\} = - 0.514$ and $\{c\sqrt{3} \Psi^2_{+1}\} = 0.134$, which yield $\{F(\mathbf{k})\}^2 \approx \{- 0.380\}^2 = 0.144$ for (1, 0, 4). These estimates are to be compared with $\{- 0.418\}^2 = 0.175$ with all odd rank multipoles, and $\{F(\mathbf{k})\}^2 = \{- 0.275\}^2 = 0.076$ with all even and odd rank multipoles. The quadrupole and hexadecapoles exert even more influence on the Bragg intensity at larger $\mathbf{k}$, as shown in Figure 2.

**4. Properties of multipoles** Steps involved in arriving at results in (3.2) merit comment [24, 25]. Let $|JMsl\rangle$ be a state with spin $s = 1/2$, orbital angular momentum l, total angular momentum J and projection M ($-J \leq M \leq J$). A matrix element of spherical tensor operator, $T^K_{Q'}$, say, satisfies the Wigner-Eckart Theorem, namely,

$$\langle JMsl| T^K_{Q'} |J'M'sl\rangle = (-1)^{K+J-J'} (2J+1)^{-1/2} (\theta||T^K||\theta') (KQ'\ J'M'|JM), \quad (4.1)$$

in which $(\theta||T^K||\theta')$ is a reduced matrix element (RME), θ and θ' are composite labels for J, s, l, and $(KQ'\ J'M'|JM)$ is a standard Clebsch-Gordan coefficient. RMEs for our two spherical tensors are written in the notation used in reference [24], with quantities $A(K'-1, K')$ and $B(K'-1, K')$ related to matrix elements of $\exp(i\mathbf{R}\cdot\mathbf{k})\ (\mathbf{k} \times \nabla)$ and $\exp(i\mathbf{R}\cdot\mathbf{k})\mathbf{s}$, respectively. We have,

$$(\theta||T^{K'}||\theta') = -(-1)^{J'-J} (2J+1)^{1/2} \{A(K'-1, K') + B(K'-1, K')\},$$

$$(\theta||T^K||\theta') = -i(-1)^{J'-J} (2J+1)^{1/2} B(K, K)$$

$$= (i^K/\sqrt{3}) (2K+1) \langle j_K \rangle (\theta||\Upsilon^K||\theta'), \quad (4.2)$$

and the operator $\Upsilon^K$ is discussed later in this section. For $A(K'-1, K')$ integer K' is odd and $K' \leq (2l-1)$, whereas in $B(K'-1, K')$ integer K' is odd with $K' \leq (2l+1)$ and $B(K, K)$ has K even and $2 \leq K \leq 2l$.

The RME of the auxiliary operator **Q** is,

$(\theta||\mathbf{Q}_p||\theta') =$

$(4\pi)^{1/2} [\Sigma_Q \Sigma_{K'Q'}\ Y^{K'-1}_Q(\kappa) ((2K'+1)/(K'+1)) (\theta||T^{K'}||\theta') (K'-1Q\ K'Q'|1p)$

$+ i \Sigma_{KQ} \Sigma_{Q'}\ Y^K_Q(\kappa) (\theta||T^K||\theta') (KQ\ KQ'|1p)], \quad (4.3)$

which should be compared to the generic expression (1.1), together with subsequent discussion. $Y^I_Q(\kappa)$ is a spherical harmonic with $\kappa = \mathbf{k}/k$. Only I even occurs in (4.3), and all $T^K$ and $T^{K'}$ are parity-even.

Returning to the quantities $B(K'-1, K')$ and $B(K, K)$, they may be expressed in terms of standard unit tensors, $W^{(1,K)K'}$ and $W^{(1,K)K}$. A key property is that $W^{(a,b)L}$ is zero for $a + b + L$ odd when electrons possess both unique J and unique l [29, 30]. In the current discussion, we have unique l (l = 2), which makes K even, and an admixture of J-states, so the aforementioned selection rules is not over-riding. The special case $W^{(1,K)K'}$ with $J = J'$ (and $K'$ odd) is explored by Ganguly et al [10].

It is also instructive to examine a natural ally of $(\theta||\exp(i\mathbf{R}\cdot\mathbf{k})\mathbf{s}||\theta')$ created by a tensor product,

$$\Upsilon^L = (-i)^{L+K+1}\{\mathbf{s} \otimes Y^K(\mathbf{n})\}^L, \qquad (4.4)$$

with $\mathbf{n} = \mathbf{R}/R$, and the phase factor makes $\Upsilon^L$ Hermitian. The RME $(l||Y^K(\mathbf{n})||l)$ vanishes for K odd. For this particular case, $(\theta||\Upsilon^L||\theta')$ is purely imaginary for $L = K$ and vanishes for $J = J'$. We find for $L = K$,

$$(\theta||\Upsilon^K||\theta') = (-i/2)(-1)^l[(2J+1)(2J'+1)/(2K+1)]^{1/2} (J\tfrac{1}{2}\ J'\tfrac{1}{2}|K1). \qquad (4.5)$$

The Clebsch-Gordan coefficient $(J\tfrac{1}{2}\ J'\tfrac{1}{2}|K\ 1) = 0$ for $J = J'$ (K even) as anticipated, and K has a minimum value 2.

Setting Q = 0 in (4.4) we derive the results, $\Upsilon^2_0 = \sqrt{(3/2)}\,[(\mathbf{s} \times \mathbf{n})_z\, n_z]$ for $K = L = 2$, and, $\Upsilon^4_0 = (\sqrt{5}/4)[(\mathbf{s} \times \mathbf{n})_z\, n_z\, (7n_z^2 - 3)]$ for $K = L = 4$. We recognize $(\mathbf{s} \times \mathbf{n})$ as the spin anapole introduced in Section 1 [22, 23]. The two expressions for $\Upsilon^K$ are manifestly time-odd and parity-even.

Lastly, $A(K'-1, K')$ is proportional to $W^{(0,K')K'}$ in which $a = 0$ as the first index occurs because $A(K'-1, K')$ arises from an operator with no spin. The rank $K'$ is odd because of a property of $\exp(i\mathbf{R}\cdot\mathbf{k})\,(\mathbf{k} \times \nabla)$, and an explicit proof is non-trivial [24].

## 5. Discussion and conclusions

A one-electron, spin-orbit-entangled ground-state derived from the $t_{2g}$ manifold of $Ir^{4+}$ ($5d^5$) orbitals that is a candidate ground-state of an iridate perovskite, $Sr_2IrO_4$, can support parity-even, time-odd quadrupoles and hexadecapoles. The strange multipoles are a product of the spin anapole and a time-even polar operator. They are absent in the $j_{eff} = 1/2$ model of iridate perovskites, because the model uses a single J-manifold of states while the strange multipoles exist only in a mixed J-manifold. Denouement of the off-quoted model will be one consequence of the unambiguous identification of the multipoles, achieved by numerical simulation of electronic structure or actual measurement.

We show, in this communication, that the strange multipoles do indeed contribute to magnetic neutron Bragg spots. In fact, they make a startling contribution to Bragg intensities, according to our numerical estimates based on a simple yet plausible model of an Ir ion (in common with the $j_{eff}$ = 1/2 model of iridate perovskites, the model used does not include quantum fluctuations or local inversion breaking). Intensities are shown to also contain contributions from triakontadipoles recently proposed as the primary order-parameter for the magnetic ground-state of $Sr_2IrO_4$ [10]. The same simulation of electronic structure is interpreted as supporting the value of the $j_{eff}$ = 1/2 model.

-----------------------------------------------------------

## Acknowledgements

We thank Professor E Balcar for an independent calculation of reduced matrix-elements. One of us (SWL) is grateful to Dr J-P Desclaux, Dr J P Clancy, Dr P Manuel, Dr D Hsieh, and Professor G van der Laan for useful correspondence and discussions.

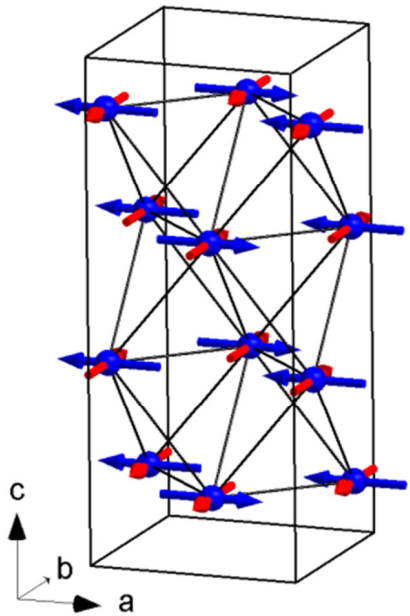

Figure 1. Motif of magnetic dipoles in $Sr_2IrO_4$ that corresponds to a bi-dimensional irreducible representation $M_4$ in a magnetic space-group $P_Icca$ ($I_pb'ca$), specified in terms of the Miller and Love notation, and Belov-Neronova-Smirnova and Opechowski-Guccione (in brackets) notations [21].

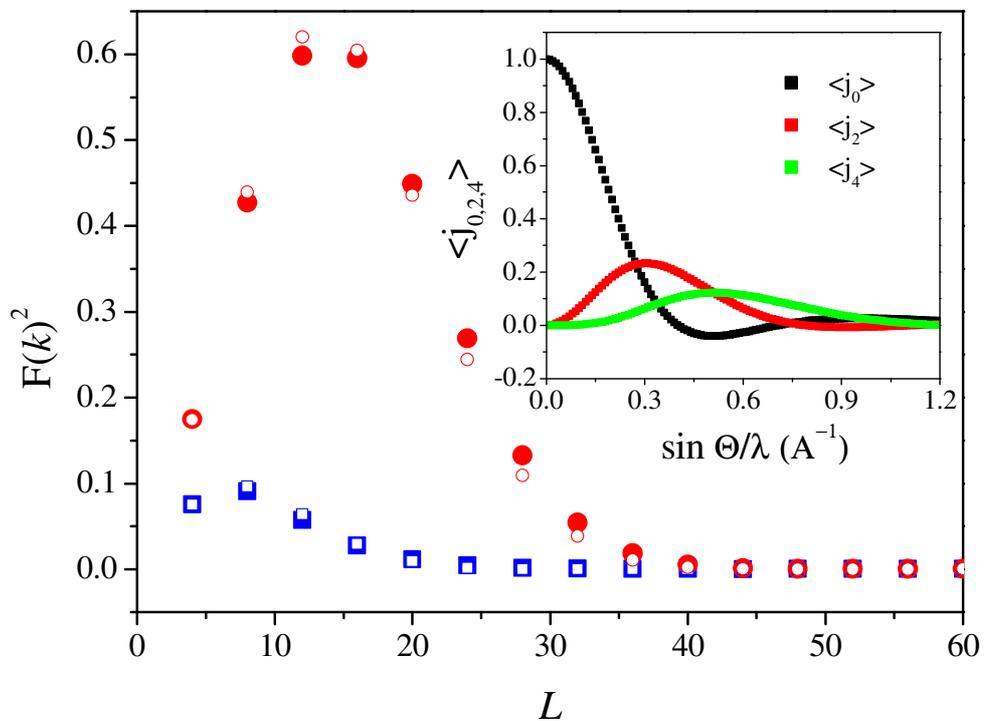

Figure 2. Intensity $\langle \mathbf{Q}_\perp \rangle \cdot \langle \mathbf{Q}_\perp \rangle = \{F(\mathbf{k})\}^2$ of the Bragg spot (1, 0, L) with L = 4n

derived from expression (2.2) using multipoles listed in (3.2) with a magnetic moment $\mu = 0.21\mu_B$, $\phi = 13.6°$ [16] and $\alpha = 0.76$, $\beta = -0.41$ and $\gamma = -0.50$. (■); expression (2.2) for F(**k**) including all multipoles of rank K (even) and K' (odd) and open symbols rank 5 (triakontadipoles) omitted. (●); multipoles K' only and open symbols rank 5 (triakontadipoles) omitted. Inset; radial integrals for $Ir^{4+}$ ($5d^5$) plotted as a function of $\sin \Theta/\lambda = |\mathbf{k}|/4\pi$, where **k** is the scattering wavevector [28].